\documentclass{article}

\usepackage[english]{babel}
\usepackage{amsmath}
\usepackage{algorithm}
\usepackage{algpseudocode}
\usepackage{amsfonts}
\usepackage{amssymb}
\usepackage[font=small,skip=0pt]{caption}
\usepackage{authblk}
\usepackage[dvipsnames]{xcolor}
\usepackage{soul}

\usepackage[letterpaper,top=2cm,bottom=2cm,left=3cm,right=3cm,marginparwidth=1.75cm]{geometry}

\usepackage{amsmath}
\usepackage{graphicx}
\usepackage[colorlinks=true, allcolors=blue]{hyperref}

\title{Phase-coded Radar Waveform Design with Quantum Annealing}

\author[1,2]{Timothé Presles}
\author[1]{Cyrille Enderli}
\author[2]{Gilles Burel}
\author[2]{El Houssaïn Baghious}
\affil[1]{Thales Defense Mission Systems, 2 av. Gay Lussac, 78851 Elancourt, France}
\affil[2]{Univ. Brest, Lab-STICC, CNRS, UMR 6285, CyberIoT Chair, 6 av. Le Gorgeu, 29200 Brest, France}

\begin{document}

\maketitle

\begin{abstract}
The Integrated Side Lobe Ratio (ISLR) problem we consider here consists in finding optimal sequences of phase shifts in order to minimize the mean squared cross-correlation side lobes of a transmitted radar signal and a mismatched replica. Currently, ISLR does not seem to be easier than the general polynomial unconstrained binary problem, which is NP-hard. In our work, we aim to take advantage of the scalability of quantum computing to find new optima, by solving the ISLR problem on a quantum annealer. This quantum device is designed to solve quadratic optimization problems with binary variables (QUBO). After proposing suitable formulation for different instances of the ISLR, we discuss the performances and the scalability of our approach on the D-Wave quantum computer. More broadly, our work enlightens the limits and potential of the adiabatic quantum computation for the solving of large instances of combinatorial optimization problems.
\end{abstract}

\section{Introduction}

Waveform optimization is an important issue in the design of radar systems. It consists in finding the "best" transmitted signal (to some criterion related to the addressed operational tasks). Under certain system constraints like e.g. peak power limitation, dissipation time, etc. The signal processing methods on receive yield undesirable side-lobes creating false alarms or masking multiple close targets. It is therefore important to mitigate these side-lobes while keeping the main-lobe energy corresponding to the actual localization of the target.

In the literature, previous works enlightens different ways to improve the quality of target or scene imaging with post-processing methods such as Spatially Variant Apodization \cite{SVA}, FFT Weighting \cite{FFT} or transmitted signal replica weighting (for mismatched filtering) \cite{replica}. In this latter case, the joint design of the transmitted signal and the replica is a more general problem of waveform optimization. \\
In this paper, we address the Integrated Side-Lobe Ratio (ISLR) problem described e.g. in \cite{Levanon}, which consists in maximizing the ratio between the energy of the main-lobe and the side-lobes. We will consider phase-coded waveforms for high range resolution applications: The problem is thus to find optimal phase codes within a finite set of quantized feasible values. Examples of contexts where this problem is important include Synthetic Aperture Radar (SAR) imagery or Ground Moving Target Indicator (GMTI) modes for airborne pulse Doppler radars. Indeed, waveforms used to implement these air-ground functions require low pulse repetition frequency and thus long pulse duration. In order to have sufficient range resolution, one has   to design intra-pulse modulation with as-low-as-possible side lobes.. In order to assess optimality, we compute the correlation function and calculate the ISLR as a performance indicator. Finding optimal sequences is NP-hard \cite{np-hard}, as the number of possible sequences exponentially increases with the length i.e. the number of possible phase shifts, and the number of phase states. In the literature, different methods are employed as branch-and-bound \cite{np-hard}, search algorithms \cite{memetic,evolutionary} or more general optimization methods \cite{opti_for_islr}. All these approaches face two major limitations: On one hand, gradient descent and continuous relaxation are very fast, but do not guarantee the quality of solutions. On the other hand, brute force search ensures optimality, but necessitates an exploration of an exponentially increasing search space, which is impossible in a reasonable time for large instances of NP-hard problems. To put it into perspective, following the results of \cite{LABS}, finding an optimal code of length $100$ (phase shifts) would take approx. 10 years of computation, and approx. $30$ trillions years for a code of length $200$. In the context of SAR imagery, code length can be as high as several thousands. \\
Quantum computing is an emerging technology which exploits the laws of quantum mechanics in order to perform logical operations. Instead of classical bits, quantum computers operate on qubits, which are in a superposition of two states. Among various applications, there is some hope that future quantum devices might provide speedup over current non-quantum approaches for solving optimization problems \cite{los_alamos}, especially for large instances of NP-Hard problems. There are currently two main approaches in the design of quantum computers: Circuit-oriented quantum computers and quantum annealers. Circuit oriented quantum computers have a sequential approach of quantum computation, using gates to perform operations on single or multiple qubits. Quantum annealers have a simultaneous approach of quantum computation, making all the qubits involved in the computation converge from an initial state to a final state. Although having, today, more qubits than their circuit oriented analogs, quantum annealers are limited to optimisation problems in the form of Quadratic Unconstrained Binary Optimisation (QUBO) problem.
QUBOs are optimisation problems with binary valued variables and a quadratic cost function which is not subject to constraints. In the so called Ising Model, the binary variables can take on values in $\{-1,1\}$. In the literature, previous work proposes numerous QUBO formulations for a wide variety of problems as clustering \cite{QUBO_cluster}, machine learning model training \cite{QUBO_ML}, graph problems \cite{QUBO_graph, QUBO_NP}...
It is worth to mention here that algorithms for solving QUBOs are also available on circuit oriented quantum computers like Quantum Approximate Optimization Algorithm (QAOA). In a previous article on the subject \cite{my_article}, we briefly studied QAOA as an alternative to annealing technologies. However, as the vast majority of algorithms developed on circuit oriented quantum computers, the error on each operators greatly limits the circuits depth i.e. the number of operators that can be used to increase the probabilities to obtain an optimal solution in the QAOA algorithm. More broadly, some authors \cite{NISQerabeyond} report that quantum computers with 50-100 qubits may be able to perform tasks which surpass the capabilities of today's classical digital computers, but noise in quantum gates will limit the size of quantum circuits that can be executed reliably.
The goal of this paper is to pursue this work by proposing a QUBO formulation for a radar waveform optimization problems, and study the performance and first results of our approach. Hence, we aim to use the scalability \cite{scalability} of quantum annealing to find new optimums for large instances of the Integrated Side-Lobe Ratio (ISLR) problem. In the following sections, we propose a QUBO formulation for the ISLR problem in its most common formulation \cite{LABS} i.e. with 2 possible phase states. Then, we extend this formulation to a discrete number of phase states. In the end of the paper, we present our first results and put them in perspective those obtained with classical approaches, then we discuss the scalability of our algorithm regarding the current state of the art of the ISLR problem and the future improvements of quantum technologies.

\section{The ISLR problem}

\subsection{The binary matched filter case}

In this part, we introduce the ISLR cost function as defined in \cite{Levanon}. For this instance, we consider that the reception filter is encoded with the same phase shifts as the emitted signal. This instance is called the matched case. We also consider only two possible values for the phases, $0$ and $\pi$, respectively associated in the following to $-1$ and $1$. Hence, the solutions take the form of a sequence of $N$ phase shift such as $S = \{s_1, \dots, s_N\}$ with $s_i \in \{-1, 1\}$. The correlation for delay $k$ is:
\begin{equation}
    C_k(S) = \sum_{i=1}^{N-k} s_is_{i+k} \label{eq:C_k} \ ,
\end{equation}
Here, due to the symmetry of the product, we only consider strictly positive values of $k$. From \eqref{eq:C_k} we define the ISLR cost function as follows:
\begin{equation}
    E(S) = \sum_{k=1}^{N-1} C_k^2(S) \label{eq:E} \ .
\end{equation}
The above equation is a degree 4 polynomial i.e. not a QUBO form. The binary matched filter case has been partially addressed in \cite{my_article}. In section 3.1, we improve the formulation by adding ancillary variables for a complete treatment of this case, resulting in a QUBO equivalent of \eqref{eq:E}. 

\subsection{The binary mismatched filter case}

In the previous section, we have defined the cost function of the binary matched case of the ISLR problem. If we now consider that the reception filter can be different, in term of length and phases shift values, from the emitted signal, we consider the mismatched case. For this formulation, we pose $S = \{s_1, \dots, s_N\}$ and $S' = \{s'_1, \dots, s'_M\}$ resp. the phases of the emitted signal and the reception filter. In that case, $S'$ can be subdivided as $S' = \{S'_A, S'_M, S'_B\}$, with $S'_A$ the prefix of length $L_A$, $S'_M$ of length $N$ and $S'_B$ the suffix of length $L_B$. Hence, we have $k \in \{-N+1, \dots, M-1\}$ and we consider the main lobe for a delay $k = L_A$. The cost function to minimize becomes:
\begin{equation}
    E_M(S,S') = \sum_{\substack{k = -N+1 \\ k \neq L_A}}^{M-1} C_k^{'2}(S,S') \ \ with \ \ C'_k(S,S') = \sum_{i=\max(1,1-k)}^{\min(N,M-k)} s_i s'_{i+k} \ . \label{eq:E_M} 
\end{equation}


\subsection{The poly-phase matched filter case}

Previous work underlines the limitation for the binary matched case in terms of ISLR minimal value. Barker codes \cite{Barker} are defined as the best possible sequences for some sequences of length $N \leq 13$, with maximal values of side-lobes equal to 1. Current state of art states that there exists no other Barker codes, at least for even sequence length \cite{Barker_even}. In order to find new optima, we can consider another instance where we no longer consider $2$ possible phase states, but a discrete number $Q$. In that case, phase states are the $Q^{th}$ roots of unity $e^{\frac{i2\pi q}{Q}}$ with $q \in \{0, \dots, Q-1\}$, and $R$ is defined as $R = \{r_1, \dots, r_N\}$, each $r_n$ being one of the $Q$ roots of unity, $n \in \{1, \dots, N\}$. In the following, we will write $\bar{z}$ the conjugate of the complex number $z$ and $\mid . \mid$ the norm operator. From \eqref{eq:C_k} and \eqref{eq:E}, we deduce the following cost function in the poly-phase matched case.
\begin{equation}
    E_Q(R) = \sum_{k=1}^{N-1} \mid \Tilde{C}_k(R) \mid^2 \ \ with \ \ \Tilde{C}_k(R) = \sum_{i=1}^{N-k} r_i\overline{r_{i+k}} \ .\label{eq:E_Q}
\end{equation}

\section{Problem implementation}

In this section, we propose an extension on our previous work presented in \cite{my_article}. We first recall some notations and results in section \ref{previous work}. Then, in section 3.2, we improve the QUBO formulation in the binary matched case. Finally, we extend this work in section 3.3 to the poly-phase case. Experimental results on quantum hardware are reported in section 4.

\subsection{Previous work} \label{previous work}

The latter proposes a QUBO formulation for the ISLR problem in the binary case. Due to the non-quadratic form of \eqref{eq:E_M}, we had to reformulate the problem in the form of a QUBO, increasing the number of variables.

In \cite{my_article}, we pose the problem as follows : Given $x \in \{-1;1\}^N$ and $y \in \{-1;1\}^M$ the sequence of spins respectively associated to the emitted sequence and the reception filter. In the following, we pose $x^T$ the transpose of the vector $x$. Then, from \eqref{eq:E_M}, we pose:
\begin{equation}
    H_C(x,y) = \sum_{k \neq L_A} (x^T Q^k y)^2 \ ,
\end{equation}
With $Q^k$ the delay matrix for a delay $k \in \{-N+1, \dots, M-1\}$ such as $Q_{i,j}^k = 1$ if $j = i + k$ and $0$ else. As mentioned in section 2.1, the ISLR cost function is not a quadratic form. Hence, we have to pose:
\begin{equation}
    z = \widetilde{xy^T} \ \ and \ \ D = \sum_{k \neq L_A} \widetilde{Q^k}\widetilde{Q^k}^T \ .
\end{equation}
Here, $\widetilde{ . }$ represents the reshaping of a $(NM)$ matrix $A$ into a $(NM,1)$ vector $a$ where $\widetilde{A_{(i,j)}} = a_{\gamma(i,j)}$ with $\gamma(i,j) = jN + i$. Straightforward calculation show that:
\begin{equation}
    H_C(z) = z^TDz \label{eq:H_C} \ .
\end{equation}
Due to this reformulation, most of $z$ solutions do not correspond to actual $x,y$ solutions. Moreover, $z$ sequences minimizing $H_C(z)$ do not correspond to actual solutions $x,y$ of $H_C(x,y)$. In order to get actual solutions, we reshape back the vector $z$ of length $NM$ into a $(N,M)$ matrix such as $Z_{i,j} = z_{\gamma(i,j)}$. If and only if $rank(Z) = 1$, we can write $Z = x y^T$ and get $x$ and $y$ respectively from the first column and row of $Z$.
In \cite{my_article}, we define the following necessary and sufficient set of conditions for $rank(Z) = 1$. Note that this set of constraints imposes that $x = y$, which corresponds to the binary matched case formulation of the ISLR problem:
\begin{enumerate}
    \item \  $\forall \ i \in \{0, \dots, N-1\}, \ Z_{i,i} = 1$ \ ,
    \item \  $\forall \ (i,j) \in \{0, \dots, N-1\}^2, \ Z_{i,j} = Z_{j,i}$ \ ,
    \item \  $\forall \ (i,j,k) \in \{0, \dots, N-1\}^3$, with $i < j < k, \ Z_{i,j}Z_{i,k} = Z_{j,k}$ \ .
\end{enumerate}
In order to guarantee that it is never profitable to not respect the constraint on the rank, we have to define a set of linear and quadratic cost functions i.e. QUBOs which are minimal if and only if $rank(Z) = 1$. 
The QUBO formulations for the first two constraints are:
\begin{equation}
    h_1(z) = N - \sum_{i=0}^{N-1} z_{\gamma(i,i)}\label {eq:h_1} \ ,
\end{equation}
\begin{equation}
    H_2(z) = \frac{(N-1)N}{2} - \sum_{\substack{i,j=0 \\ i < j}}^{N-1} z_{\gamma(i,j)}z_{\gamma(j,i)} \label{eq:H_2} \ .
\end{equation}
For the sake of clarity in the following section, we add constants to these formulations in order to set their minimal values to zero. However, during the optimization, these constants can be removed without impacting the quality of the result.

\subsection{Improvements on QUBO formulation for the binary matched case}

In \cite{my_article}, we concluded that the third constraint, due to its non-quadratic formulation, could not be straightforward implemented as a QUBO. In \cite{non_qubo_formulation}, it is demonstrated that there exists no quadratic equivalent of the third constraint without addition of ancillary variables. Hence, we propose a QUBO formulation for the third constraint which requires a set of $N_\alpha = \binom{N}{3}$ ancillary variables $z^\alpha \in \{-1,1\}^{N_\alpha}$, inspired by the work presented in \cite{non_qubo_formulation}. For the sake of simplicity, we consider that $z^\alpha_{(i,j,k)}$ is the ancillary variable associated to the triplet $(i,j,k)$ where $0 \leq i < j < k \leq N$.
\begin{equation}
    H_3(z,z^\alpha) = 4\binom{N}{3} + H_{3Q}(z,z^\alpha) + h_{3L}(z,z^\alpha) \ , \label{eq:H_3}
\end{equation}
\begin{align}
    H_{3Q}(z,z^\alpha) = \sum_{\substack{i,j,k = 0 \\ i < j < k}}^{N-1} & (  z_{\gamma(i,j)}z_{\gamma(i,k)} - z_{\gamma(i,j)}z_{\gamma(j,k)} -  z_{\gamma(i,k)}z_{\gamma(j,k)} \dots \nonumber \\
    & \dots -2 z_{\gamma(i,j)}z^\alpha_{(i,j,k)} \ ,
    -2 z_{\gamma(i,k)}z^\alpha_{(i,j,k)} 
    +2 z_{\gamma(j,k)}z^\alpha_{(i,j,k)} ) \ .
\end{align}

\begin{equation}
    h_{3L}(z,z^\alpha) = \sum_{\substack{i,j,k = 0 \\ i < j < k}}^{N-1} \left( z_{\gamma(i,j)} + z_{\gamma(i,k)} - z_{\gamma(j,k)} -2z^\alpha_{(i,j,k)} \right) \ .
\end{equation}
From \eqref{eq:H_C}, \eqref{eq:h_1}, \eqref{eq:H_2} and \eqref{eq:H_3}, we deduce the QUBO formulation for solving the ISLR problem in the binary matched case :
\begin{equation}
    H(z, z^\alpha) = H_C(z) + \lambda_1h_1(z) + \lambda_2H_2(z) + \lambda_3H_3(z,z^\alpha)  \ . \label{eq:H_final}
\end{equation}
With $\lambda_1$, $\lambda_2$ and $\lambda_3$ positive real factors related to Lagrange multipliers. In order to prevent that a non-viable solution minimizes \eqref{eq:H_final}, we have to define minimal values for the Lagrange multipliers.
First, straightforward calculations from \eqref{eq:E} show that $\smash{\displaystyle\max_{S_{ _{ }}}} \ E(S) = \frac{N(N-1)(2N-1)}{6}$. In the following, we set $\Delta_N = \smash{\displaystyle\max_{S_{ _{ }}}} \ E(S)$ as the difference between the lowest and the highest possible value of $H(z)$. As the non-respect of constraints could lead to $H_C(z) = 0$, there is no strictly positive lower bound for the ISLR criteria in this case. However, in the literature, lower bounds for \eqref{eq:E} can be found in the binary matched case \cite{Levanon}. \\
For $\lambda_1$, lets consider the worst case scenario where the optimal solution of $H(z)$ implies that a single diagonal value of $Z$ is equal to $-1$. In that situation, $h_1(z) = 2$, and we have to set $\lambda_1 > \frac{\Delta_N}{2}$ in order to guarantee that it is never profitable to break the constraint to get a better solution. \\
For $\lambda_2$, we consider a similar scenario where for a single couple of $(i,j)$, $Z_{i,j} \neq Z_{j,i}$. In this situation, $H_2(z) = 2$ and we have to set $\lambda_2 > \frac{\Delta_N}{2}$. For $\lambda_3$, we also consider a scenario where the constraint is respected except for a single triplet $(i,j,k)$. Hence, $H_3(z) \geq 2$ and we have to set $\lambda_3 > \frac{\Delta_N}{2}$. \\
In practice, these Lagrange multiplies can be set at lower positive values. We consider here that a single non-respect of constraint will reduce the value of $H_C(z)$ from the worst to the best possible solution, which is not the case for most instances with different values of $N$.

\subsection{QUBO formulation for the polyphase matched case}

In this section, we propose a QUBO formulation of \eqref{eq:E_Q} in the polyphase matched case. In this formulation, we consider a phase code of length $N$. Recall that each phase state can take $Q$ possible values defined as the $Q^{th}$ roots of unity $e^{\frac{i2\pi q}{Q}}$. In order to implement these complex and discrete phase state, we pose $b = \{0,1\}^{NQ}$. $b$ is composed of $N$ blocs of length $Q$ such as each bloc $b_i$ has exactly one non-null value at index $q \in \{0, \dots, Q-1\}$. We also pose the $(N,NQ)$ phase matrix:
\begin{equation}
    \Phi = 
    \begin{pmatrix}
        [e^{0}, \dots, e^{\frac{i2\pi (Q-1)}{Q}}] & \dots & 0 \\
        \vdots & \ddots & \vdots \\
        0 & \dots & [e^{0}, \dots, e^{\frac{i2\pi (Q-1)}{Q}}]
    \end{pmatrix} \ .
\end{equation}
Straightforward calculations show that $x=\Phi b$, with $x \in \{e^{\frac{i2\pi q}{Q}}, q = 0, \dots, Q-1 \}^N$ encoding the complex phase values. With the same approach as in section 3.1, we pose :
\begin{equation}
    \begin{split}
           & A^k = \Phi^*Q^k\Phi \ , \\
           & D' = \sum_{k \neq 0} \widetilde{A^k} \widetilde{A^{k^T}}^T \ , \\
           & z' = \widetilde{bb^T} \label{eq:reform} \ .
    \end{split}
\end{equation}
In \eqref{eq:reform}, $\Phi^*$ is the conjugate transpose $\overline{\Phi^T}$ of $\Phi$. Straightforward calculations show that:
\begin{equation}
    H'_C(z) = z^{\prime T}D' z' \label{eq:H'_C}  \ .
\end{equation}
As a consequence of the reformulation, most of $z'$ solutions do not correspond to actual $b$ solutions. As in 3.1, in order to get actual solutions, $z'$ of length $(NQ)^2$ has to be reshaped into a $(NQ,NQ)$ matrix $Z'$. $Z'$ is composed of $N^2$ blocs $Z^{(n,m)}, n,m \in \{0, \dots ,N-1\}^2$ of shape $(Q,Q)$ having a single non-null element each such as:
\begin{equation}
    Z' = 
    \begin{pmatrix}
        Z'^{(0,0)} & \dots & Z'^{(0,N-1)} \\
        \vdots & \ddots & \vdots \\
        Z'^{(N-1,0)} & \dots & Z'^{(N-1,N-1)}
    \end{pmatrix}  \ .
\end{equation}

In the following, we will note $Z'^{(n,m)}_{i,j}$ the term at the $i^{th}$ row and $j^{th}$ column of the bloc $Z'^{(n,m)}$. We pose $\phi(n,m) = nNQ + mQ$ and $\mu(n,m,i,j) = \phi(n,m) + \gamma(i,j)$ such as $z'_{\mu(n,m,i,j)} = Z'^{(n,m)}_{i,j}$ with $n,m \in \{0,\dots, N-1\}^2$ and $i,j \in \{0,\dots, Q-1\}^2$. 
In order to guarantee that it is never favorable, in term of value of the cost function, to not respect the one-hot encoding, we pose the following constraint cost function :
\begin{equation}
    H'_{OH}(z') = \sum_{n=0}^{N-1} \sum_{m=0}^{N-1} \left( \sum_{i=0}^{Q-1} \sum_{j=0}^{Q-1} z'_{\mu(n,m,i,j)} - 1\right)^2 \label{eq:H'_OH} \ .
\end{equation}
As in section 3.1, we can write $Z' = bb^T$ and get the values of $x$ if and only if $rank(Z') = 1$. Then, by admitting that the one-hot encoding is respected, we define the following set of necessary and sufficient conditions $\forall n,m \in \{0, \dots, N-1\}^2$ and $\forall i,j \in \{0, \dots, Q-1\}^2$:
\begin{enumerate}
    \item \  $Z'^{(n,m)}_{i,j} = Z'^{(m,n)}_{j,i}$ \ ,
    \item \  If $Z'^{(n,n)}_{i,i} = 1$, then $\sum_{m=0}^{N-1} \sum_{j=0}^{Q-1} Z'^{(m,n)}_{j,i} = N$ and $\sum_{m=0}^{N-1} \sum_{j=0}^{Q-1} Z'^{(n,m)}_{i,j} = N$ \ ,
    \item \  If $Z'^{(n,n)}_{i,i} = 0$, then $\sum_{m=0}^{N-1} \sum_{j=0}^{Q-1} Z'^{(m,n)}_{j,i} = 0$ and $\sum_{m=0}^{N-1} \sum_{j=0}^{Q-1} Z'^{(n,m)}_{i,j} = 0$ \ .
\end{enumerate}

\begin{figure}[h!]
    \centering
    \captionsetup{justification=centering,margin=2cm}
    \includegraphics[scale=0.6]{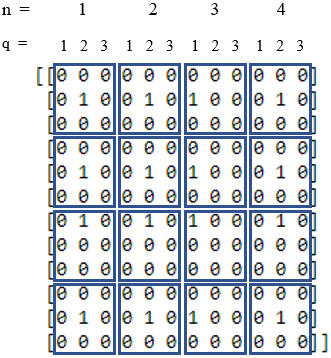}
    \caption{Example of solution matrix $Z$ for $N = 4$ and $Q = 3$. The value of each phase state $n$ is retrieved by looking at the index $q$ of each non-null value in the first row of blocs. Here, the corresponding sequence is $x = [e^{\frac{i2\pi}{3}}, e^{\frac{i2\pi}{3}}, e^{0}, e^{\frac{i2\pi}{3}}]$}
    \label{fig:example_form}
\end{figure}

With $Z'$ taking values in $\{0,1\}$, the second and third conditions imply that for given $p,q$, if $Z'^{(p,q)}_{i,j} = 1$, then $\forall n,m \in \{0, \dots, N-1\}^2$, $Z'^{(n,m)}_{i,j} = 1$. This results in a $Z'$ matrix full of zeros except for $N$ rows (resp. columns) having exactly $N$ non-null values which implies that $rank(Z') = 1$. For the first condition, we pose the following cost function inspired by \eqref{eq:H_2} :
\begin{equation}
    H'_{S}(z') = \sum_{n=0}^{N-1} \sum_{n<m}^{N-1} \sum_{i=0}^{Q-1} \sum_{i<j}^{Q-1} z'_{\mu(n,m,i,j)} + z'_{\mu(m,n,j,i)} - 2 z'_{\mu(n,m,i,j)} z'_{\mu(m,n,j,i)}  \ .\label{eq:H'_S}
\end{equation}

Straightforward calculation from the second condition leads to a $3^{rd}$ degree polynomial, which is not a QUBO form. However, as one-hot encoding is imposed, we know that each row (resp. column) of each block $Z'^{(n,m)}_{i,j}$ contains at most $1$ non-null value. This implies that if the non-null value is at a given row (resp. column) of the block, it cannot be at another row (resp. column) of the block. If we impose that all rows (resp. columns) of each horizontally (resp. vertically) adjacent blocks have to be equal, then for a given $n,m \in \{0, \dots, N-2\}^2$ and $i$ (resp. $j$) $\in \{0, \dots, Q-1\}$:
\begin{equation}
    \begin{split}
           & \sum_{j=0}^{Q-1} Z'^{(n,m)}_{i,j} = \sum_{j=0}^{Q-1} Z'^{(n,m+1)}_{i,j} \ , \\
           \text{resp.} \ \ & \sum_{i=0}^{Q-1} Z'^{(n,m)}_{i,j} = \sum_{i=0}^{Q-1} Z'^{(n+1,m)}_{i,j} \ . \label{eq:just}
    \end{split}
\end{equation}
Note that the above equalities both guarantee the second and third conditions, as the value of the sums, equal to $0$ or $1$, imply that the sum of values of each rows of $Z'$ is equal to $0$ or $N$. Hence, from \eqref{eq:just}, straightforward calculation allow us to define the constraint cost function for both the second and the third conditions :
\begin{equation}
    H_{CH}(z') = \sum_{n=0}^{N-1} \sum_{i=0}^{Q-1} \sum_{m=0}^{N-2} \left( \sum_{j=0}^{Q-1} z'_{\mu(n,m,i,j)} - \sum_{j=0}^{Q-1} z'_{\mu(n,m+1,i,j)} \right)^2  \ , \label{eq:H'_CH}
\end{equation}
\begin{equation}
    H_{CV}(z') = \sum_{m=0}^{N-1} \sum_{j=0}^{Q-1} \sum_{n=0}^{N-2} \left( \sum_{i=0}^{Q-1} z'_{\mu(n,m,i,j)} - \sum_{i=0}^{Q-1} z'_{\mu(n+1,m,i,j)} \right)^2 \label{eq:H'_CV}  \ .
\end{equation}

From \eqref{eq:H'_C}, \eqref{eq:H'_OH}, \eqref{eq:H'_S}, \eqref{eq:H'_CV} and \eqref{eq:H'_CH}, we deduce the following cost function for solving the ISLR problem in the $Q$-phase states matched case :
\begin{equation}
    H'(z') = H_{C}(z') + \lambda_{OH}H'_{OH}(z') + \lambda_SH'_{S}(z') + \lambda_{CH}H_{CH}(z') + \lambda_{CV}H_{CV}(z') \label{eq:H'} \ .
\end{equation}
With $\lambda_{OH}$, $\lambda_{S}$, $\lambda_{CH}$ and $\lambda_{CV}$ positive real factors called Lagrange multipliers. As in \eqref{eq:H_final}, we have to define minimal values for these multipliers by considering the worst case scenario favoring the non respect of their associated constraints. Then, we can pose $\Delta_{N,Q}$ as in section 3.2, which corresponds to the difference between the best and the worst possible value, and define the Lagrange multipliers following this value. For $\lambda_{OH}$, a single non-respect of the corresponding constraint implies that $H_{OH}(z') = 1$ which leads to setting $\lambda_{OH} > \Delta_{N,Q}$. Same goes for $\lambda_{S}$ and $\lambda_{CH}$ as respectively $H'_{S}(z') = 1$ and $H'_{CH}(z') = 1$ in case of a single non respect, which leads to $\lambda_{S} > \Delta_{N,Q}$ and $\lambda_{CH} > \Delta_{N,Q}$. Due to the similar form of $H'_{CV}(z')$ and $H'_{CH}(z')$, we can set $\lambda_{CV} = \lambda_{CH}$. As for the binary case, in practice, these Lagrange multipliers can be set at lower values, as near-optimal values of $H'_C(z')$ can be obtained with respect of all the constraints. 



\section{Results}

\subsection{Experimental context}

\begin{figure}[h!]
    \centering
    \captionsetup{justification=centering,margin=2cm}
    \includegraphics[scale=0.55]{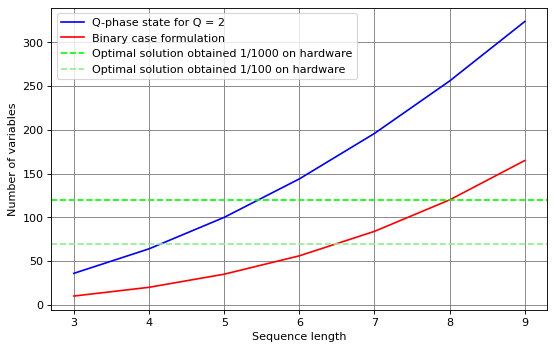}
    \caption{Number of variables required to implement the binary (red slope) and poly-phase (blue slope) formulations for small instances of the ISLR problem}
    \label{fig:obtention_time_vs_nb_variables}
\end{figure}

In this section, we propose and discuss the results obtained on the D-Wave adiabatic quantum computer \cite{D-Wave}. D-Wave is a Canadian company which proposes access to their quantum devices through a cloud based service called D-Wave Leap. The device we use in the following is the D-Wave Advantage quantum computer, having 5760 available qubits and 15 couplers per qubits. The number of couplers quantifies the number of interconnections available between qubits i.e. variables. In the QUBO formulations presented above, a product between two variables implies there is a coupler between their two corresponding qubits. As this number of products can exceed 15, the embedding function provided by D-Wave libraries binds qubits with each other, creating logical qubits from multiple physical qubits. In the computation, logical qubits act as a single variable. Consequently, due to this embedding constraint, the actual limitation of problem size that can be implemented on the quantum machine is below the theoretical limitation ($(NQ)^2 < 5760$ for example for the Q-states formulation). Recent publications from D-Wave present their hardware having 7800+ qubits available and more interconnections. However, in the following, we will consider current limitations of D-Wave hardware at 5760 qubits and 15 interconnections available. \\
In order to compare our results, we also used a simulated annealing library provided by D-Wave called dwave-neal. First, we used this tool in order to implement larger instances of the problem which cannot be implemented in the quantum hardware. Then, due to the inherent noise of quantum computation, multiple executions i.e. samples were required in order to guarantee the quality of solutions obtained and correctly estimate the average computation time for a given instance of the problem. \textbf{Fig. \ref{fig:obtention_time_vs_nb_variables}} represents the number of variables as a function of the sequence length $N$. Green dotted lines represent the estimated number of variables for which one has $1/100$ or $1/1000$ probability of obtaining an optimal solution. Nevertheless, note that actual solutions (respecting all constraints without minimizing the ISLR criteria) are more often obtained. Unfortunately, as we write these lines, dwave-neal does not provide a quantum noise simulator. Consequently, our estimations of the time required to obtain a solution for any instance will be based on an extrapolation of computation time for small instances.

In the following, we define the computation time for a single sample as a sum of the delay time, the annealing time and the readout time. Delay time is fixed at $21 \mu s$, and consist in preparing the system in an equiprobability of obtention of any sequence. Then, the annealing time represents the convergence time from the initial state to the final state. The latter is parameterized by the QUBO, and its minimal energy configurations correspond to solutions minimizing the QUBO i.e. solutions respecting all constraints and minimizing the ISLR criteria. Previous work shown that the optimal annealing time logarithmically scales with the number of variables \cite{temps_annealing}. For this problem, search heuristics have shown that optimal solutions were most probably obtained for annealing time values between $50 \mu s$ and $250 \mu s$ with a number of variables scaling from $10$ to $144$. The readout time also logarithmically scales with the number of variables, from $80 \mu s$ to $115 \mu s$ for variables scaling from $10$ to $144$.

\subsection{Results in the binary matched case}

\begin{figure}[h!]
    \centering
    \captionsetup{justification=centering,margin=2cm}
    \includegraphics[scale=0.53]{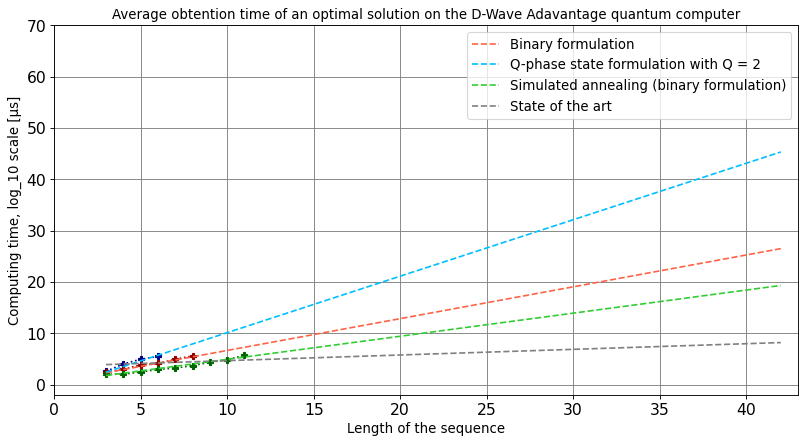}
    \caption{Comparison of computation time for different instances of the ISLR problem and classical approaches in the binary matched case. Points represent the average computation time required to obtain an optimal solution on D-Wave hardware. Corresponding dotted lines are an extrapolation of the computation time for larger instances. The grey dotted line represents the average computation time with classical approaches following \cite{LABS}.}
    \label{fig:time_islr_bin}
\end{figure}

As a reminder, this section will consider submitting the binary formulation defined by \eqref{eq:H_final} and the $Q$-phase states formulation defined by \eqref{eq:H'} with $Q=2$ to the D-Wave quantum computer. For this instance of the ISLR problem, current limitations in terms of sequence length i.e. the value of $N$ are presented in various papers \cite{ISLR_hard1, ISLR_hard2, ISLR_hard3}. Our first results are presented in \textbf{Fig. \ref{fig:time_islr_bin}} for both binary and poly-phase state formulations. On this graph, computation time is obtained by multiplying the average number of samples required to obtain an optimal solution by the computation time of a single sample. For the simulated annealing plot, the average number of samples is obtained on a classical simulator, but the computation time is based on the quantum annealing obtention time. The goal of this plot is to represent the obtention time on a quantum computer with similar error rates than simulated annealing. Finally, state-of-art plot corresponds to results presented in \cite{ISLR_classical} with different classical approaches.

For these results and for both formulations, we have set all the Lagrange multipliers at $\lambda = 10$. As discussed in section 3.2, these multipliers can be set as lower values than the theory (based on worst case scenario) in order to increase the probability of success (obtaining a global minimum). Indeed, as the quantum annealer uses tunneling effect in order to converge to a global minima, decreasing the energy of the potential barriers between configurations increases the probability of the system to cross the barrier and reach a new minima \cite{D-Wave}. Search heuristics shown that on the D-Wave hardware we used for our first results, high values of Lagrange multipliers reduces the number of optimal solutions. This is due to the relative low difference in energy between sub-optimal actual solutions (minimizing constraint cost functions without minimizing \eqref{eq:H_C}) and optimal solutions.

\subsection{Results in the poly-phase states case}

In this section, we present the improvements in the value of the cost function for a discrete number of phase state. In the previous section, we presented a straightforward correlation between the computation time and the number of variables. Hence, obtention time for the $Q$-phase state instances can be deduced from the obtention time in the binary case presented in sections 4.1 and 4.2, as the number of logical variables required is $(NQ)^2$, with $N$ the length of the sequence and $Q$ the number of phase states. However, for small $N$ and $Q$ values, simulations and brute-force computation provided new minimum for the ISLR criteria, presented in Fig. 3. These results underline the advantage of increasing the number of phase states, especially for high values of N. 

\begin{figure}[h!]
    \centering
    \captionsetup{justification=centering,margin=2cm}
    \includegraphics[scale=0.6]{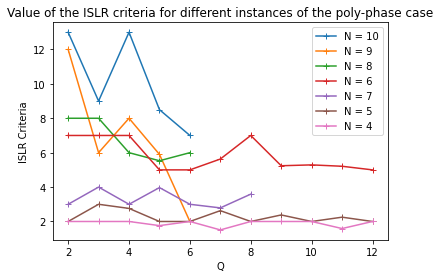}
    \caption{Value of the ISLR criteria for Q $\geq$ 2 phase states}
    \label{fig:q_state_res}
\end{figure}

In the literature, \cite{Levanon} statistically demonstrated that for high values of $N$ and when $N=Q$, poly-phase Barker codes can be found. Poly-phase Barker codes have a maximal peak-to-side-lobe (PSL) value of 1, which implies that the ISLR criteria is at most equal to $N$. These results can also be compared to periodic poly-phases codes like Golomb sequences \cite{Levanon} for example.

\subsection{Comparisons between the approaches and limits of the model}

In section 3, we presented two approaches to solve the ISLR problem in binary and poly-phase cases. Due to the quadratic reformulation, especially in the binary case, size of the search space has been increased. Consequently, we restricted our results to the matched case, where the reception filter has the same phase code as the emitted signal, in order to limit the size of the search space and present our first results. Moreover, the mismatched case \eqref{eq:E_M} requires an additional condition to maximize the value of the main-lobe, which also has to be implemented as a penalty cost function in the form of a QUBO. In our present work, we have not found any way in the literature yet to implement it, or at least without greatly increasing the size of the search space. Nevertheless, note that removing the symmetry cost function in \eqref{eq:H'} extends the formulation to the mismatched case, without any guarantee of optimality for the main-lobe value.\\
Disregarding the error of the quantum hardware, highlighted in \textbf{Fig. \ref{fig:time_islr_bin}}, larger instances may be implemented by duplicating some variables with a lot of connections. Even if D-Wave embedding functions already binds logical variables to multiple physical qubits, these generic functions may not provide an optimal embedding. Hence, including the optimal embedding in the QUBO formulation could increase the number of implementable instances, at the cost of additional ancillary variables. \\
In term of scalability in the binary matched case, for $N > 21$, the number of variables required to encode the binary case formulation $N^2 + \binom{N}{3}$ is greater that the number of variables required for the $Q$-phase state formulation $4N^2$. Moreover, optimal sequences of length $N < 21$ have already been found. Hence, in term of number of variables, the $Q$-phase state formulation with $Q = 2$ would be more profitable than the binary case formulation in order to find new optimums for the ISLR problem. More precisely, with this formulation, approx. $25000$ logical qubits would be required to address instances only approximately solved in the state-of-the-art ($N \approx 80$) that currently take $\approx 40$ days to be solved by classical approaches following \cite{LABS}. \\


\section{Conclusion}

In conclusion, our work proposes a new method to find optimal waveforms in the binary and poly-phase matched case, which is compatible with quantum annealing technology of D-Wave. We also present our first results on the D-Wave quantum computer and discuss the scalability of our algorithm regarding the machine hardware constraints. We conclude that current limitations are hardware based, and that improvements of the machine would close the gap between our results and optimal phase sequences for large instances. In the poly-phase state, our first results underlines the advantage of generalizing the formulation to a discrete number of phases, by proposing new minimums for the ISLR criteria. \\
To complete this paper, future work will consist in studying the optimal parameters of the model, especially for the annealing time and the way the problem is embedded. It could be also interesting to test our method on other quantum annealers, as the Pasqal machine \cite{pasqal}, which uses spatially arranged cold atoms to perform quantum annealing. For all these technologies, studies could also be done on the total computation overhead, including the embedding time and limitations due to hardware architectures and qubit technologies, namely spatial arrangement of cold atoms for Pasqal. Moreover, we could find QUBO formulations to optimize other signal processing criteria as Doppler tolerance \cite{Doppler} or the PSL criteria \cite{PSL}.

\bibliographystyle{IEEEtran}

\bibliography{sample}

\begin{thebibliography}{10}
\providecommand{\url}[1]{#1}
\csname url@samestyle\endcsname
\providecommand{\newblock}{\relax}
\providecommand{\bibinfo}[2]{#2}
\providecommand{\BIBentrySTDinterwordspacing}{\spaceskip=0pt\relax}
\providecommand{\BIBentryALTinterwordstretchfactor}{4}
\providecommand{\BIBentryALTinterwordspacing}{\spaceskip=\fontdimen2\font plus
\BIBentryALTinterwordstretchfactor\fontdimen3\font minus
  \fontdimen4\font\relax}
\providecommand{\BIBforeignlanguage}[2]{{%
\expandafter\ifx\csname l@#1\endcsname\relax
\typeout{** WARNING: IEEEtran.bst: No hyphenation pattern has been}%
\typeout{** loaded for the language `#1'. Using the pattern for}%
\typeout{** the default language instead.}%
\else
\language=\csname l@#1\endcsname
\fi
#2}}
\providecommand{\BIBdecl}{\relax}
\BIBdecl

\bibitem{SVA}
C.~F. Castillo-Rubio, S.~Llorente-Romano, and M.~Burgos-Garcia, ``Spatially
  variant apodization for squinted synthetic aperture radar images,''
  \emph{IEEE Transactions on Image Processing}, vol.~16, no.~8, pp. 2023--2027,
  2007.

\bibitem{FFT}
F.~Arlery, R.~Kassab, U.~Tan, and F.~Lehmann, ``Efficient optimization of the
  ambiguity functions of multi-static radar waveforms,'' in \emph{2016 17th
  International Radar Symposium (IRS)}, 2016, pp. 1--6.

\bibitem{replica}
F.~Xin, B.~Wang, S.~Li, X.~Song, and C.-H. Wang, ``Adaptive radar waveform
  design based on weighted mi and the difference of two mutual information
  metrics,'' \emph{Complexity}, vol. 2021, pp. 1--12, 01 2021.

\bibitem{Levanon}
\BIBentryALTinterwordspacing
``Matched filter,'' pp. 20--33, 2004. [Online]. Available:
  \url{https://onlinelibrary.wiley.com/doi/abs/10.1002/0471663085.ch2}
\BIBentrySTDinterwordspacing

\bibitem{np-hard}
T.~Packebusch and S.~Mertens, ``Low autocorrelation binary sequences,''
  \emph{Journal of Physics A: Mathematical and Theoretical}, vol.~49, 12 2015.

\bibitem{memetic}
J.~Gallardo, C.~Cotta, and A.~Fernández-Leiva, ``Finding low autocorrelation
  binary sequences with memetic algorithms,'' \emph{Applied Soft Computing},
  vol.~9, pp. 1252--1262, 09 2009.

\bibitem{evolutionary}
B.~Militzer, M.~Zamparelli, and D.~Beule, ``Evolutionary search for low
  autocorrelated binary sequences,'' \emph{IEEE Transactions on Evolutionary
  Computation}, vol.~2, no.~1, pp. 34--39, 1998.

\bibitem{opti_for_islr}
\BIBentryALTinterwordspacing
U.~Tan, O.~Rabaste, C.~Adnet, F.~Arlery, and J.-P. Ovarlez, ``{Optimization
  Methods for Solving the Low Autocorrelation Sidelobes Problem},'' pp. 1--5,
  May 2016. [Online]. Available: \url{https://hal.science/hal-01360260}
\BIBentrySTDinterwordspacing

\bibitem{LABS}
\BIBentryALTinterwordspacing
T.~Packebusch and S.~Mertens, ``Low autocorrelation binary sequences,''
  \emph{Journal of Physics A: Mathematical and Theoretical}, vol.~49, no.~16,
  p. 165001, mar 2016. [Online]. Available:
  \url{https://doi.org/10.1088%2F1751-8113%2F49%2F16%2F165001}
\BIBentrySTDinterwordspacing

\bibitem{los_alamos}
B.~Tasseff, T.~Albash, Z.~Morrell, M.~Vuffray, A.~Y. Lokhov, S.~Misra, and
  C.~Coffrin, ``On the emerging potential of quantum annealing hardware for
  combinatorial optimization,'' 2022.

\bibitem{QUBO_cluster}
C.~Bauckhage, N.~Piatkowski, R.~Sifa, D.~Hecker, and S.~Wrobel, ``A qubo
  formulation of the k-medoids problem,'' 09 2019.

\bibitem{QUBO_ML}
\BIBentryALTinterwordspacing
P.~Date, D.~Arthur, and L.~Pusey-Nazzaro, ``{QUBO} formulations for training
  machine learning models,'' \emph{Scientific Reports}, vol.~11, no.~1, may
  2021. [Online]. Available: \url{https://doi.org/10.1038%2Fs41598-021-89461-4}
\BIBentrySTDinterwordspacing

\bibitem{QUBO_graph}
C.~Calude, M.~Dinneen, and R.~Hua, ``Qubo formulations for the graph
  isomorphism problem and related problems,'' \emph{Theoretical Computer
  Science}, vol. 701, 06 2017.

\bibitem{QUBO_NP}
\BIBentryALTinterwordspacing
A.~Lucas, ``Ising formulations of many {NP} problems,'' \emph{Frontiers in
  Physics}, vol.~2, 2014. [Online]. Available:
  \url{https://doi.org/10.3389%2Ffphy.2014.00005}
\BIBentrySTDinterwordspacing

\bibitem{my_article}
\BIBentryALTinterwordspacing
T.~Presles, C.~Enderli, R.~Bricout, F.~Aligne, and F.~Barba-Resco,
  ``{Phase-coded radar waveform AI-based augmented engineering and optimal
  design by Quantum Annealing},'' Aug. 2021, working paper or preprint.
  [Online]. Available: \url{https://hal.science/hal-03318130}
\BIBentrySTDinterwordspacing

\bibitem{NISQerabeyond}
\BIBentryALTinterwordspacing
J.~Preskill, ``Quantum {C}omputing in the {NISQ} era and beyond,''
  \emph{{Quantum}}, vol.~2, p.~79, Aug. 2018. [Online]. Available:
  \url{https://doi.org/10.22331/q-2018-08-06-79}
\BIBentrySTDinterwordspacing

\bibitem{scalability}
L.~Gyongyosi and S.~Imre, ``A survey on quantum computing technology,''
  \emph{Computer Science Review}, vol.~31, pp. 51--71, 02 2019.

\bibitem{Barker}
M.~Trevorrow and D.~Farmer, ``The use of barker codes in doppler sonar
  measurements,'' \emph{Journal of Atmospheric and Oceanic Technology - J ATMOS
  OCEAN TECHNOL}, vol.~9, pp. 699--704, 10 1992.

\bibitem{Barker_even}
J.~Willms, ``A note on barker sequences of even length,'' 2021.

\bibitem{non_qubo_formulation}
A.~Mandal, A.~Roy, S.~Upadhyay, and H.~Ushijima-Mwesigwa, ``Compressed
  quadratization of higher order binary optimization problems,'' 2020.

\bibitem{D-Wave}
M.~Johnson, M.~Amin, S.~Gildert, T.~Lanting, F.~Hamze, N.~Dickson, R.~Harris,
  A.~Berkley, J.~Johansson, P.~Bunyk, E.~Chapple, C.~Enderud, J.~Hilton,
  K.~Karimi, E.~Ladizinsky, N.~Ladizinsky, T.~Oh, I.~Perminov, C.~Rich, and
  G.~Rose, ``Quantum annealing with manufactured spins,'' \emph{Nature}, vol.
  473, pp. 194--8, 05 2011.

\bibitem{temps_annealing}
O.~Galindo and V.~Kreinovich, ``What is the optimal annealing schedule in
  quantum annealing,'' pp. 963--967, 2020.

\bibitem{ISLR_hard1}
E.~Rodríguez-Heck, ``Linear and quadratic reformulations of nonlinear
  optimization problems in binary variables,'' \emph{4OR}, vol.~17, 11 2018.

\bibitem{ISLR_hard2}
J.~Gallardo, C.~Cotta, and A.~Fernández-Leiva, ``Finding low autocorrelation
  binary sequences with memetic algorithms,'' \emph{Applied Soft Computing},
  vol.~9, pp. 1252--1262, 09 2009.

\bibitem{ISLR_hard3}
D.~Padilha, ``Solving np-hard problems on an adiabatic quantum computer,''
  Ph.D. dissertation, 10 2014.

\bibitem{ISLR_classical}
M.~V. H~Houdrouge, ``Low auto-correlation binary sequences,'' 8 2019.

\bibitem{pasqal}
\BIBentryALTinterwordspacing
``Pasqal quantum computer.'' [Online]. Available: \url{https://www.pasqal.com/}
\BIBentrySTDinterwordspacing

\bibitem{Doppler}
R.~Amar, M.~Alaee-Kerahroodi, P.~Babu, and B.~S.~M. R., ``Designing
  interference-immune doppler-tolerant waveforms for radar systems,''
  \emph{IEEE Transactions on Aerospace and Electronic Systems}, pp. 1--20,
  2022.

\bibitem{PSL}
M.~F. Keskin, R.~F. Tigrek, C.~Aydogdu, F.~Lampel, H.~Wymeersch, A.~Alvarado,
  and F.~M. Willems, ``Peak sidelobe level based waveform optimization for ofdm
  joint radar-communications,'' pp. 1--4, 2021.

\end{thebibliography}

\end{document}